\documentclass[conference]{IEEEtran}
\usepackage{amsmath,amsfonts}
\usepackage{graphicx,amssymb}
\usepackage{algorithmic,algorithm}
\usepackage{amsthm,dsfont}
\usepackage{subfigure,url}
\usepackage[noadjust]{cite}

\theoremstyle{plain}
\newtheorem{lem}{Lemma}

\newtheorem{prop}[]{Proposition}

\newcounter{longequ}[longequ]
\addtocounter{longequ}{11}

\newcommand\relphantom[1]{\mathrel{\phantom{#1}}}
\allowdisplaybreaks	

\newcommand{\Nt}{{N_\mathrm{t}}}

\IEEEoverridecommandlockouts
\begin{document}
%
\title{Enabling Secure Wireless Communications via Intelligent Reflecting Surfaces}
\author{\IEEEauthorblockN{Xianghao Yu, Dongfang Xu, and Robert Schober}
\IEEEauthorblockA{Friedrich-Alexander-Universit\"{a}t Erlangen-N\"{u}rnberg, Germany\\
Email: \{xianghao.yu, dongfang.xu, robert.schober\}@fau.de
}
}


%


\maketitle

\begin{abstract}
In this paper, we propose to utilize intelligent reflecting surfaces (IRSs) for enhancing the physical layer security of wireless communications systems.
In particular, an IRS-assisted secure wireless system is considered, where a multi-antenna transmitter communicates with a single-antenna receiver in the presence of an eavesdropper. 
To maximize the secrecy rate, both the  beamformer at the transmitter and the IRS phase shifts are jointly optimized. 
Based on the block coordinate descent (BCD) and minorization maximization (MM) techniques, two efficient algorithms are developed to solve the resulting non-convex optimization problem for small- and large-scale IRSs, respectively.
Simulation results show that IRSs can significantly improve physical layer security if the proposed algorithms are employed. Furthermore, we reveal that deploying large-scale IRSs is more efficient than enlarging the antenna array size of the transmitter for both boosting the secrecy rate and enhancing the energy efficiency.
\end{abstract}

\IEEEpeerreviewmaketitle

\section{Introduction}
Physical layer security has received considerable attention from both academia and industry in recent years \cite{5751298}. Various approaches to improve physical layer security have been proposed in the literature, e.g., cooperative relaying schemes, artificial noise-aided beamforming \cite{8349956}, and cooperative jamming \cite{5352243}. However, deploying a large number of relays or other helpers in secure wireless systems inevitably incurs an excessive cost. Furthermore, cooperative jamming and transmitting artificial noise consume additional power for security provisioning.  These shortcomings of existing approaches urgently call for a new paradigm for secure wireless systems, which is both cost-effective and energy-efficient.

With the rapid development of radio frequency (RF) micro-electro-mechanical systems (MEMS), programmable and reconfigurable meta-surfaces have recently found abundant applications in the public and civil domains, among which intelligent reflecting surfaces (IRSs) have drawn special attention for their applications in wireless communications \cite{di2019smart}. The artificial thin films of IRSs can be easily coated on existing infrastructures such as walls of buildings, which reduces implementation cost and complexity. In addition, unlike conventional communication transceivers, IRSs consume no power as they are passive devices. Furthermore, different from conventional transceivers, IRSs help transmit information without  generating new signals. Instead, they smartly transform or recycle existing signals. To sum up, IRSs are passive cost-effective devices with the ability to control the radio propagation environment \cite{8466374}. These characteristics make IRSs promising key enablers for improving the physical layer security of wireless communications in an economical and energy-efficient manner.

There are several studies on the design of IRS-assisted wireless systems \cite{8855810,8811733,Wu_discrete,nadeem2019large,huang2018large}. A point-to-point multiple-input single-output (MISO) system was investigated in \cite{8811733,Wu_discrete}, where the IRS is implemented with continuous and discrete phase shifters, respectively. 
Based on the semidefinite relaxation (SDR) method, approximate solutions for the beamformer at the access point and the phase shifts at the IRS were developed. 
The authors of \cite{nadeem2019large} considered a downlink multiuser communication system, where the signal-to-interference-plus-noise ratio (SINR) was maximized for given phase shifts. Energy efficiency maximization was tackled in \cite{huang2018large}, where sub-optimum zero-forcing beamforming was assumed at the access point. We note that none of these existing works considers physical layer security, despite its great importance for modern wireless systems.

To fill this gap, this paper investigates physical layer security provisioning for IRS-assisted wireless systems. Assume a transmitter equipped with multiple antennas communicates with one legitimate receiver in the presence of an eavesdropper. Both the legitimate receiver and the eavesdropper are assumed to use a single antenna, and the IRS is implemented via programmable phase shifters. Our goal is to maximize the secrecy rate of the considered system by optimizing both the beamformer at the transmitter and the phase shifts at the IRS, which leads to a non-convex optimization problem. 
Based on the block coordinate descent (BCD) and  minorization maximization (MM) techniques, two efficient algorithms are proposed for solving the problem. 
The first algorithm is more suitable for small-scale IRSs, while the second algorithm is advantageous for  large-scale IRSs. Unlike existing works \cite{8811733,nadeem2019large,huang2018large}, we obtain locally optimal solutions for both the beamformer and the phase shifts. 
To the best of the authors' knowledge, this is the first work that studies the design of secure IRS-assisted wireless systems.

\emph{Notations:} The imaginary unit of a complex number is denoted by $\jmath=\sqrt{-1}$.
Matrices and vectors are denoted by boldface
capital and lower-case letters, respectively. $\mathbb{C}^{m\times n}$
denotes the set of all $m\times n$ complex-valued
matrices. $\mathbf{I}_m$ is the $m$-dimensional identity matrix. The $i$-th element of vector $\mathbf{a}$ is denoted as $a_i$. $\mathbf{A}^*$ and $\mathbf{A}^H$ stand for the conjugate and conjugate transpose of matrix $\mathbf{A}$. 
$\mathrm{diag}(a_1,\cdots , a_n)$ denotes
a diagonal matrix whose diagonal entries are $a_1,\cdots, a_n$.
The largest eigenvalue of matrix $\mathbf{A}$ and the corresponding eigenvector are denoted by $\lambda_{\max}(\mathbf{A})$ and $\boldsymbol{\lambda}_{\max}(\mathbf{A})$, respectively.
$\triangleq$ means ``defined as''.
Expectation and the real part of a complex number are denoted by $\mathbb{E}[\cdot]$ and $\Re(\cdot)$, respectively. The operation $\angle(\mathbf{A})$ constructs a matrix by extracting the phases of the elements of matrix $\mathbf{A}$.
\begin{figure}[t]
	\centering\includegraphics[width=6cm]{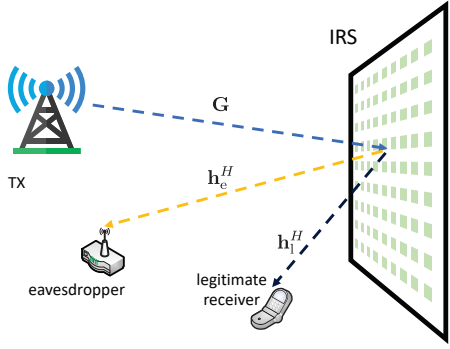}
	\caption{An IRS-assisted secure communication system.
	}\label{model}
\end{figure}
\section{System Model}
Consider an IRS-assisted  communication system, which consists of a transmitter, one legitimate receiver, an eavesdropper, and an IRS, as shown in Fig. \ref{model}. We assume that the transmitter is equipped with $\Nt$ antennas, while the legitimate receiver and the eavesdropper use a single receive antenna, respectively. The passive IRS is deployed in the network to improve the physical layer security, and employs $M$ phase shifters. 
Equipped with a controller, 
the phase shifts of the IRS are programmable. Furthermore, we assume a quasi-static flat-fading channel model and perfect channel state information (CSI) knowledge at both the transmitter and the IRS\footnote{While the CSI of the eavesdropper is generally difficult to acquire, the results in this paper serve as theoretical performance upper bounds for the considered system. These bounds and the insights gained from them can be used to guide the system design for the case when the CSI of the eavesdropper is not perfectly known.}. 
The received baseband signals at the legitimate receiver and the eavesdropper can be expressed as
\begin{equation}
\begin{split}
&y_\mathrm{l}=\mathbf{h}^H_\mathrm{l}\mathbf{\Phi}\mathbf{G}\mathbf{f}x+n_\mathrm{l},\\
&y_\mathrm{e}=\mathbf{h}^H_\mathrm{e}\mathbf{\Phi}\mathbf{G}\mathbf{f}x+n_\mathrm{e},\\
\end{split}
\end{equation}
where $\mathbf{h}_\mathrm{l}\in\mathbb{C}^{M\times1}$ and $\mathbf{h}_\mathrm{e}\in\mathbb{C}^{M\times1}$ represent the channels from the IRS to the legitimate receiver and eavesdropper, respectively. 
The phase shift matrix $\mathbf{\Phi}$  of the IRS is given by $\mathbf{\Phi}=\mathrm{diag}(e^{\jmath \theta_1},e^{\jmath \theta_2},\cdots,e^{\jmath \theta_M})$, where $\theta_k$ is the phase shift of the $k$-th reflecting element of the IRS \cite{8811733}.
The channel matrix from the transmitter to the IRS is denoted as $\mathbf{G}\in\mathbb{C}^{M\times\Nt}$, and the linear beamforming vector at the transmitter side is denoted as $\mathbf{f}\in\mathbb{C}^{\Nt\times1}$. The  signal transmitted to the legitimate receiver is denoted as $x$, where $\mathbb{E}[|x|^2]=1$ without loss of generality. 
$n_\mathrm{l}$ and $n_\mathrm{e}$ are additive complex Gaussian noises with variances $\sigma_\mathrm{l}^2$ and $\sigma_\mathrm{e}^2$, respectively.

The achievable secrecy rate of the IRS-assisted MISO wireless system is given by \cite{5961840}
\begin{equation}\label{eq2}
\left[\log\left(1+\frac{1}{\sigma_\mathrm{l}^2}\left|\mathbf{h}_\mathrm{l}^H\mathbf{\Phi Gf}\right|^2\right)-
\log\left(1+\frac{1}{\sigma_\mathrm{e}^2}\left|\mathbf{h}_\mathrm{e}^H\mathbf{\Phi Gf}\right|^2\right)\right]^+,
\end{equation}
where $[x]^+=\max\{0,x\}$.
Our goal in this paper is to maximize the secrecy rate by optimizing the transmit beamforming vector $\mathbf{f}$ and the phase shift matrix $\mathbf{\Phi}$ of the IRS. We note that dropping the operator $[\cdot]^+$ in \eqref{eq2} has no impact on the optimization\footnote{
	The secrecy rate is zero if the transmission is turned off. Hence, the term $\log\left(1+\frac{1}{\sigma_\mathrm{l}^2}\left|\mathbf{h}_\mathrm{l}^H\mathbf{\Phi Gf}\right|^2\right)-\log\left(1+\frac{1}{\sigma_\mathrm{e}^2}\left|\mathbf{h}_\mathrm{e}^H\mathbf{\Phi Gf}\right|^2\right)$ will always be non-negative if the beamformer and the phase shifts are optimized to maximize the secrecy rate.
	}. The resulting optimization is formulated as
\begin{equation}\mathcal{P}_1:\,
\begin{aligned}
&\underset{\mathbf{f},\mathbf{\Phi}}{\mathrm{maximize}} && \frac{1+\frac{1}{\sigma_\mathrm{l}^2}\left|\mathbf{h}_\mathrm{l}^H\mathbf{\Phi}\mathbf{G}\mathbf{f}\right|^2}{1+\frac{1}{\sigma_\mathrm{e}^2}\left|\mathbf{h}_\mathrm{e}^H\mathbf{\Phi}\mathbf{G}\mathbf{f}\right|^2}\\
&\mathrm{subject\thinspace to}&&\left\Vert\mathbf{f}\right\Vert^2\le P\\
&&&\mathbf{\Phi}=\mathrm{diag}\left(e^{\jmath \theta_1},e^{\jmath \theta_2},\cdots,e^{\jmath \theta_M}\right),
\end{aligned}
\end{equation}
where $P\ge0$ is the given total transmit power.

\emph{Remark 1:} Note that, different from secure wireless systems without IRSs, in the second constraint of optimization problem $\mathcal{P}_1$, each diagonal element in the phase shift matrix $\mathbf{\Phi}$ has unit modulus, i.e., $|e^{\jmath\theta_k}|=1$. This non-convex constraint together with the non-convex objective function makes $\mathcal{P}_1$ a non-convex problem. The globally optimal solution of non-convex optimization problems with unit modulus constraints is in general not tractable \cite{NIPS2013_5041}.


\begin{figure*}
		\newcounter{TempEqCnt}                         			
	\setcounter{TempEqCnt}{\value{equation}} 			
	\setcounter{equation}{\value{longequ}}          	
	\begin{equation}\label{longeq}
	\tilde{\theta}_k=
	\arctan\frac{c_{\mathrm{e},k}d_{\mathrm{l},k}\cos\left(p_{\mathrm{l},k}\right) - c_{\mathrm{l},k}d_{\mathrm{e},k}\cos\left(p_{\mathrm{e},k}\right)}
	{c_{\mathrm{e},k}d_{\mathrm{l},k}\sin\left(p_{\mathrm{l},k}\right) - c_{\mathrm{l},k}d_{\mathrm{e},k}\sin\left(p_{\mathrm{e},k}\right)}-\arccos\frac{d_{\mathrm{l},k}d_{\mathrm{e},k}\sin\left(p_{\mathrm{e},k}-p_{\mathrm{l},k}\right)}
	{\sqrt{c_{\mathrm{e},k}^2d_{\mathrm{l},k}^2+c_{\mathrm{l},k}^2d_{\mathrm{e},k}^2-2c_{\mathrm{l},k}c_{\mathrm{e},k}d_{\mathrm{l},k}d_{\mathrm{e},k}\cos\left(p_{\mathrm{l},k}-p_{\mathrm{e},k}\right)}}
	\end{equation}\hrule
\end{figure*}
\setcounter{equation}{\value{TempEqCnt}} 		


\section{Design of Secure IRS-Assisted Wireless Systems}
Block coordinate descent (BCD) methods optimize the objective function with respect to different subsets (blocks) of optimization variables in each iteration while the other blocks are fixed. 
BCD has been shown to be  a widely applicable and empirically successful approach in many applications \cite{NIPS2013_5041,7397861}, and typically leads to a sub-optimal solution for non-convex problems.
In this paper, we resort to this approach as the main methodology for solving $\mathcal{P}_1$ efficiently.
\subsection{Transmit Beamformer Design}
According to the principle of BCD, we first investigate the optimization of beamforming vector $\mathbf{f}$ for a fixed phase shift matrix $\mathbf{\Phi}$. The beamformer design problem is accordingly given by 
\begin{equation}\mathcal{P}_2:\,
\underset{||{\mathbf{f}}||^2\le P}{\mathrm{maximize}}\quad\frac{1+\frac{1}{\sigma_\mathrm{l}^2}\left|\mathbf{h}_\mathrm{l}^H\mathbf{\Phi}\mathbf{G}\mathbf{f}\right|^2}{1+\frac{1}{\sigma_\mathrm{e}^2}\left|\mathbf{h}_\mathrm{e}^H\mathbf{\Phi}\mathbf{G}\mathbf{f}\right|^2},
\end{equation}
and the optimal solution is provided in the following lemma.
\begin{lem} \label{lem1}
Given the phase shift matrix $\mathbf{\Phi}$ of the IRS, the optimal solution for beamforming vector $\mathbf{f}$  is given by
\begin{equation}\label{eq4}
\mathbf{f}^\star=\sqrt{P}\boldsymbol{\lambda}_{\max}\left(\mathbf{X}_\mathrm{e}^{-1}\mathbf{X}_l\right),
\end{equation}	
where
\begin{equation}
\mathbf{X}_i=\mathbf{I}_{N_\mathrm{t}}+\frac{P}{\sigma_i^2}\mathbf{G}^H\mathbf{\Phi}^H\mathbf{h}_{i}\mathbf{h}_{i}^H\mathbf{\Phi}\mathbf{G},\quad i\in\{\mathrm{l,e}\}.
\end{equation}
\end{lem}
\begin{IEEEproof}
	It was shown in \cite{4557589} that allocating all transmit power for beamforming is optimal, i.e., $\left\Vert\mathbf{f}^\star\right\Vert^2=P$. Then, the numerator and denominator of the objective function of $\mathcal{P}_2$ can be rewritten as
	\begin{align}
	1+\frac{P}{\sigma_i^2}\left|\mathbf{h}_{i}^H\mathbf{\Phi}\mathbf{G}\tilde{\mathbf{f}}\right|^2&=\tilde{\mathbf{f}}^H\tilde{\mathbf{f}}+\frac{P}{\sigma_i^2}\tilde{\mathbf{f}}^H\left(\mathbf{G}^H\mathbf{\Phi}^H\mathbf{h}_{i}\mathbf{h}_{i}^H\mathbf{\Phi}\mathbf{G}\right)\tilde{\mathbf{f}}\nonumber\\
	&\triangleq\tilde{\mathbf{f}}^H\mathbf{X}_i\tilde{\mathbf{f}},\label{eq6}
	\end{align}
	where $\tilde{\mathbf{f}}=\mathbf{f}/\sqrt{P}$ is a unit vector. By substituting \eqref{eq6} into the objection function of $\mathcal{P}_2$, we can rewrite $\mathcal{P}_2$ as
	\begin{equation}
	\underset{||\tilde{\mathbf{f}}||^2=1}{\mathrm{maximize}}\quad\frac{\tilde{\mathbf{f}}^H\mathbf{X}_\mathrm{l}\tilde{\mathbf{f}}}
	{\tilde{\mathbf{f}}^H\mathbf{X}_\mathrm{e}\tilde{\mathbf{f}}}.
	\end{equation}
	In this way, we transform $\mathcal{P}_2$ to a generalized eigenvalue problem, whose optimal solution is given by \eqref{eq4}.
\end{IEEEproof}
\emph{Remark 2:} The result in Lemma 1 is similar to secure beamforming design for MISO communications without IRSs, for which a closed-form solution is available \cite{4557589}. In particular, the beamformer $\mathbf{f}$ is  designed to be as orthogonal to the effective eavesdropping channel $\mathbf{h}_\mathrm{e}^H\mathbf{\Phi}\mathbf{G}$ as possible, while being as aligned with the effective legitimate receiver channel $\mathbf{h}_\mathrm{l}^H\mathbf{\Phi}\mathbf{G}$ as possible. Compared to conventional secure communications systems, the incorporation of the IRS adds another degree of freedom (DoF) to establish favorable effective channels $\mathbf{h}_\mathrm{l}^H\mathbf{\Phi}\mathbf{G}$ and $\mathbf{h}_\mathrm{e}^H\mathbf{\Phi}\mathbf{G}$ by carefully choosing the phase shift matrix $\mathbf{\Phi}$.

To the best of the  authors' knowledge, there is no general approach for the optimal design of the phase shift matrix $\mathbf{\Phi}$. Hence, in the following two subsections, we propose two different approaches for optimizing $\mathbf{\Phi}$ in the BCD procedure.

\subsection{Element-Wise BCD}

In this subsection, we adopt an element-wise BCD for optimizing the phase shift matrix $\mathbf{\Phi}$. In other words, we take each phase shift $\theta_k$ as one block in the BCD. The corresponding optimization problem is  given by
\begin{equation}\label{eq9}\mathcal{P}_3:\,
\begin{aligned}
&\underset{\theta_k}{\mathrm{maximize}} && \frac{1+\frac{1}{\sigma_\mathrm{l}^2}\left|\mathbf{h}_\mathrm{l}^H\mathbf{\Phi}\mathbf{G}\mathbf{f}\right|^2}{1+\frac{1}{\sigma_\mathrm{e}^2}\left|\mathbf{h}_\mathrm{e}^H\mathbf{\Phi}\mathbf{G}\mathbf{f}\right|^2}\\
&\mathrm{subject\thinspace to}&&\mathbf{\Phi}=\mathrm{diag}\left(e^{\jmath \theta_1},\cdots,e^{\jmath \theta_k},\cdots,e^{\jmath \theta_M}\right).
\end{aligned}
\end{equation}
The optimal solution is presented in the following lemma.
\begin{lem}\label{lem2}
	Given the beamforming vector $\mathbf{f}$ and phase shifts $\{\theta_m\}_{m\ne k}$, the optimal solution for $\theta_k$ is given by
	\begin{equation}\label{eq10}
	\theta_k^\star=
	\begin{cases}
	\tilde{\theta}_k+\pi&c_{\mathrm{e},k}d_{\mathrm{l},k}\cos\left(p_{\mathrm{l},k}\right) < c_{\mathrm{l},k}d_{\mathrm{e},k}\cos\left(p_{\mathrm{e},k}\right)\\
	\tilde{\theta}_k&\text{otherwise},
	\end{cases}
	\end{equation}
	where 
	\begin{align}\label{eq11}
	c_{i,k}&=\frac{1}{2}\left(1+\frac{1}{\sigma_i^2}\left|h_{i,k}^*\mathbf{g}_k^H\mathbf{f}\right|^2
					+\frac{1}{\sigma_i^2}\left|\sum_{m\ne k}h_{i,m}^*e^{\jmath\theta_m}\mathbf{g}_m^H\mathbf{f}\right|^2\right),\nonumber\\
	 d_{i,k}&=\frac{1}{\sigma_i^2}\left|h_{i,k}^*\mathbf{g}_k^H\mathbf{f}\sum_{m\ne k}h_{i,m}e^{-\jmath\theta_m}\mathbf{f}^H\mathbf{g}_m\right|, \\
	  p_{i,k}&=\angle\left(h_{i,k}^*\mathbf{g}_k^H\mathbf{f}\sum_{m\ne k}h_{i,m}e^{-\jmath\theta_m}\mathbf{f}^H\mathbf{g}_m\right),\quad i\in\{\mathrm{l,e}\},\nonumber
	\end{align}
	$\mathbf{g}_k^H$ is the $k$-th row of matrix $\mathbf{G}$,
	and $\tilde{\theta}_k$ is given by \eqref{longeq} shown on top of this page.
\end{lem}
\begin{IEEEproof}
	See Appendix \ref{appA}.
\end{IEEEproof}
\setcounter{equation}{12} 	

\begin{algorithm}[t]
	\caption{Element-Wise BCD}
	\begin{algorithmic}[1]
		\STATE Construct an initial $\mathbf{\Phi}^{(0)}$ and set $t=0$;
		\REPEAT 
		\STATE Fix $\mathbf{\Phi}^{(t)}$ and optimize $\mathbf{f}^{(t)}$ according to \eqref{eq4};
		\FOR{$k=1$ \TO $M$ } \STATE {Optimize the phase shift $\theta_k^{(t+1)}$ according to \eqref{eq10};} \ENDFOR
		\STATE $t\leftarrow t+1$;
		\UNTIL convergence;
	\end{algorithmic}
\end{algorithm}

With Lemmas \ref{lem1} and \ref{lem2} at hand, the element-wise BCD is summarized in \textbf{Algorithm 1}. As the closed-form globally optimal solutions in  \eqref{eq4} and \eqref{eq10} are used in each block of the element-wise BCD, the objective function monotonically increases. In addition,  it is easy to verify that the objective function is upper bounded by the point-to-point MISO channel capacity. These two properties together guarantee that \textbf{Algorithm 1} converges to a locally optimal solution of $\mathcal{P}_1$. However, the number of BCD blocks is $M+1$ since each phase shift is a block. This leads to a slow convergence for large IRS sizes $M$. 

\subsection{Alternating Optimization With MM}
Instead of treating each phase $\theta_k$ as a single block in the BCD, in this subsection, we take the entire phase shift matrix $\mathbf{\Phi}$ as one block. Consequently, there are only two blocks in the BCD. Hence, the BCD reduces to the special case of alternating optimization (AO).
By leveraging the minorization maximization (MM) technique, we update all phase shifts $\{\theta_k\}_{k=1}^M$ in parallel in each iteration.

We reformulate the optimization of the phase shift matrix $\mathbf{\Phi}$ as follows. Using
\begin{equation}\label{eq13}
\mathbf{h}_{i}^H\mathbf{\Phi}\mathbf{G}=\mathbf{v}^H\mathbf{R}_i,\quad i\in\{\mathrm{l,e}\},
\end{equation}
where $\mathbf{v}=[e^{\jmath\theta_1},\dots,e^{\jmath\theta_M}]^H$ and $\mathbf{R}_i=\mathrm{diag}(\mathbf{h}^H_{i})\mathbf{G}$, 
the numerator and denominator in \eqref{eq9} can be rewritten as
\begin{equation}
\begin{split}
1+\frac{1}{\sigma_i^2}\left|\mathbf{h}_{i}^H\mathbf{\Phi}\mathbf{G}\mathbf{f}\right|^2&\overset{(a)}{=}\frac{1}{M}\mathbf{v}^H\mathbf{v}+\frac{1}{\sigma_i^2}\mathbf{v}^H\mathbf{R}_i\mathbf{ff}^H\mathbf{R}_i^H\mathbf{v}\\
&\,\triangleq\mathbf{v}^H\mathbf{Y}_i\mathbf{v},
\end{split}
\end{equation}
where  $\mathbf{Y}_i=\frac{1}{M}\mathbf{I}_M+\frac{P}{\sigma_i^2}\mathbf{R}_i\mathbf{ff}^H\mathbf{R}_i^H$, and step $(a)$ exploits $\mathbf{v}^H\mathbf{v}=M$.
Therefore,  optimization problem $\mathcal{P}_3$ can be recast as
\begin{equation}\mathcal{P}_4:\,
\begin{aligned}
&\underset{\mathbf{v}}{\mathrm{maximize}} && g(\mathbf{v})= \frac{\mathbf{v}^H\mathbf{Y}_\mathrm{l}\mathbf{v}}{\mathbf{v}^H\mathbf{Y}_\mathrm{e}\mathbf{v}}\\
&\mathrm{subject\thinspace to}&&|v_k|=1,\quad k\in\{1,2,\cdots,M\}.
\end{aligned}
\end{equation}

\emph{Remark 3:} The main difficulty in solving $\mathcal{P}_4$ is the unit modulus constraint, which is element-wise and highly non-convex. By introducing $\mathbf{V}=\mathbf{vv}^H$ as the optimization variable and relaxing the rank-one constraint of $\mathbf{V}$, $\mathcal{P}_4$ can be solved via  semidefinite relaxation (SDR). The resulting problem is a quasi-convex problem, whose optimal solution can be obtained by solving a series of semidefinite programming (SDP) problems.
However, there is no guarantee that the obtained solution $\mathbf{V}$ is a rank-one matrix, and therefore the SDR approach can only provide an approximate solution for $\mathbf{v}$. 
As a result, the objective function does not necessarily increase in each iteration of the AO, and the convergence of the SDR-based algorithm cannot be guaranteed. In addition, the computational complexity of solving a large number of SDP problems in each iteration of the AO is prohibitively high.

\begin{figure}[t]
	\centering\includegraphics[width=7cm]{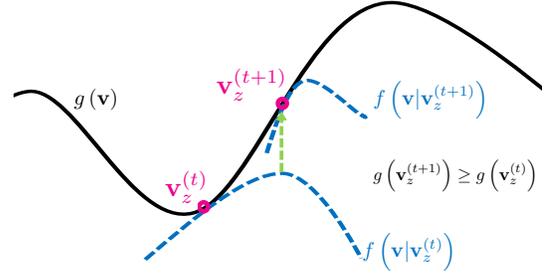}
	\caption{The procedure of minorization maximization. 
	}\label{MM}
\end{figure}

In this paper, we propose to solve $\mathcal{P}_4$ by the MM technique \cite{7547360}, whose main idea is illustrated in Fig. \ref{MM}. In particular, assuming the value of $\mathbf{v}$ in the $t$-th iteration of the AO is denoted as $\mathbf{v}_z^{(t)}$, we construct a lower bound on the objective function $g(\mathbf{v})$ that touches the objective function at  point $\mathbf{v}_z^{(t)}$, denoted as $f\left(\mathbf{v}|\mathbf{v}_z^{(t)}\right)$. We adopt this lower bound as a surrogate objective function, and the maximizer of this surrogate objective function is then taken as the value of $\mathbf{v}$ in the next iteration of the AO, i.e., $\mathbf{v}_z^{(t+1)}$. In this way, the objective value is monotonically increasing from one iteration to the next, i.e., $g\left(\mathbf{v}_z^{(t+1)}\right)\ge
g\left(\mathbf{v}_z^{(t)}\right)$. The key to the success of MM lies in constructing a surrogate objective function $f\left(\mathbf{v}|\mathbf{v}_z^{(t)}\right)$ for which the maximizer $\mathbf{v}_z^{(t+1)}$ is easy to find. For phase shift matrix optimization problem $\mathcal{P}_4$, a surrogate objective function is composed in the following lemma.

\begin{lem}\label{lem3}
	The objective function $g(\mathbf{v})$ is lower bounded by
	\begin{equation}
	\begin{split}
	g(\mathbf{v})&=\frac{\mathbf{v}^H\mathbf{Y}_\mathrm{l}\mathbf{v}}{\mathbf{v}^H\mathbf{Y}_\mathrm{e}\mathbf{v}}\ge f\left(\mathbf{v}|\mathbf{v}_z\right)+\left[g(\mathbf{v}_z)-f(\mathbf{v}_z|\mathbf{v}_z)\right],\\
	\end{split}
	\end{equation}
	where
	\begin{align}
	f\left(\mathbf{v}|\mathbf{v}_z\right)&=
	2\frac{\Re\left(\mathbf{v}_z^H\mathbf{Y}_\mathrm{l}\mathbf{v}\right)}{\mathbf{v}_z^H\mathbf{Y}_\mathrm{e}\mathbf{v}_z}-\frac{\mathbf{v}_z^H\mathbf{Y}_\mathrm{l}\mathbf{v}_z}{\left(\mathbf{v}_z^H\mathbf{Y}_\mathrm{e}\mathbf{v}_z\right)^2}\Big\{\mathbf{v}^H\lambda_{\max}(\mathbf{Y}_\mathrm{e})\mathbf{v}\nonumber\\
	&\relphantom{=}+2\Re\left(\mathbf{v}_z^H\left[\mathbf{\mathbf{Y}_\mathrm{e}}-\lambda_{\max}(\mathbf{Y}_\mathrm{e})\mathbf{I}_M\right]\mathbf{v}\right)\Big\},\label{eq17}
	\end{align}
	and $g(\mathbf{v}_z)-f(\mathbf{v}_z|\mathbf{v}_z)$ is a constant term that is irrelevant for optimization.
\end{lem}
\begin{IEEEproof}
	Defining $y=\mathbf{v}^H\mathbf{Y}_\mathrm{e}\mathbf{v}$, the objective function $g(\mathbf{v})\triangleq\frac{\mathbf{v}^H\mathbf{Y}_\mathrm{l}\mathbf{v}}{y}$ is jointly convex in $\{\mathbf{v},y\}$ since $\mathbf{Y}_\mathrm{l}=\frac{1}{M}\mathbf{I}_M+\frac{P}{\sigma_\mathrm{l}^2}\mathbf{R}_\mathrm{l}\mathbf{ff}^H\mathbf{R}_\mathrm{l}^H$ is positive definite. Because of the convexity, we have the following inequality
	\begin{equation}
	\begin{split}
	\frac{\mathbf{v}^H\mathbf{Y}_\mathrm{l}\mathbf{v}}{\mathbf{v}^H\mathbf{Y}_\mathrm{e}\mathbf{v}}
	&\ge2\frac{\Re\left(\mathbf{v}_z^H\mathbf{Y}_\mathrm{l}\mathbf{v}\right)}{\mathbf{v}_z^H\mathbf{Y}_\mathrm{e}\mathbf{v}_z}-\frac{\mathbf{v}_z^H\mathbf{Y}_\mathrm{l}\mathbf{v}_z}{\left(\mathbf{v}_z^H\mathbf{Y}_\mathrm{e}\mathbf{v}_z\right)^2}\mathbf{v}^H\mathbf{Y}_\mathrm{e}\mathbf{v}+c\\
	&\overset{(b)}{\ge} f\left(\mathbf{v}|\mathbf{v}_z\right)+\left[g(\mathbf{v}_z)-f(\mathbf{v}_z|\mathbf{v}_z)\right],
	\end{split}
	\end{equation}where $c$ is a constant and $(b)$ applies \cite[Lemma 1]{7093191}.
\end{IEEEproof}
\begin{prop}
	The phase shift optimization problem in each iteration of the AO is equivalent to
\begin{equation}
		\mathcal{P}_5:\,\mathbf{v}_z^{(t+1)}=\arg\underset{|v_i|=1}{\max}\quad\Re\left[\left(\mathbf{w}^{(t)}\right)^H\mathbf{v}\right],
\end{equation}
where
\begin{equation}\label{eq20}
\begin{split}
\mathbf{w}^{(t)}=\frac{\mathbf{Y}_\mathrm{l}\mathbf{v}_z^{(t)}}{\left(\mathbf{v}_z^{(t)}\right)^H\mathbf{Y}_\mathrm{e}\mathbf{v}_z^{(t)}}-\frac{\left(\mathbf{v}_z^{(t)}\right)^H\mathbf{Y}_\mathrm{l}\mathbf{v}_z^{(t)}}{\left[\left(\mathbf{v}_z^{(t)}\right)^H\mathbf{Y}_\mathrm{e}\mathbf{v}_z^{(t)}\right]^2}\\
\times\left[\mathbf{\mathbf{Y}_\mathrm{e}}-\lambda_{\max}(\mathbf{Y}_\mathrm{e})\mathbf{I}_M\right]\mathbf{v}_z^{(t)}.
\end{split}
\end{equation}
The optimal solution of $\mathcal{P}_5$ is given by\footnote{Since $\mathbf{v}_z$ is a unit modulus vector, it is sufficient to determine the phases of the elements of the vector.}
\begin{equation}\label{eq21}
\angle\left(\mathbf{v}_z^{(t+1)}\right)=\angle\left(\mathbf{w}^{(t)}\right).
\end{equation}
\end{prop}
\begin{IEEEproof}
	With Lemma \ref{lem3}, the optimization problem that needs to be solved is the maximization of $f\left(\mathbf{v}|\mathbf{v}_z\right)$ in \eqref{eq17} with respect to $\mathbf{v}$. Since $\mathbf{v}^H\mathbf{v}$ is a constant, the proposition can be proved with basic algebraic manipulations.
\end{IEEEproof}

The AO-MM algorithm is presented in \textbf{Algorithm 2}. With the closed-form solutions in \eqref{eq4} and \eqref{eq21}, the objective function is guaranteed to monotonically increase and to converge to a local optimum.
\begin{algorithm}[t]
	\caption{Alternating Optimization With Minorization Maximization (AO-MM)}
	\begin{algorithmic}[1]
		\STATE Construct an initial $\mathbf{v}_z^{(0)}$ and set $t=0$;
		\REPEAT 
		\STATE Fix $\mathbf{v}_z^{(t)}$ and optimize $\mathbf{f}^{(t)}$ according to \eqref{eq4};
		\STATE Optimize $\mathbf{v}_z^{(t+1)}$ according to  \eqref{eq20} and \eqref{eq21};
		\STATE $t\leftarrow t+1$;
		\UNTIL convergence.
	\end{algorithmic}
\end{algorithm}
There are some open issues that require further remarks.

\emph{(1) Initial point: }In both the element-wise BCD and AO-MM algorithms, we require an initialization for the phase shifts of the IRS. As $\mathcal{P}_1$ is a non-convex problem, the quality of the sub-optimal solution depends to some extent  on the initialization. Here, we propose an effective way to construct the initial values of the phase shifts. In particular, we set
\begin{equation}
\angle\left(\mathbf{v}_z^{(0)}\right)=\angle\left(\mathbf{u}\right),\quad \mathbf{\Phi}^{(0)}=\mathrm{diag}\left(\mathbf{v}_z^{(0)}\right),
\end{equation}
where $\mathbf{u}$ is the dominant left singular vector of $\mathbf{R}_\mathrm{l}$ in \eqref{eq13}. This initialization is a heuristic obtained by ignoring the denominator of the objective function.

\emph{(2) Complexity comparison:} As closed-form solutions are obtained for all the blocks of both algorithms, the computational complexity of the proposed algorithms is critically determined by the total number of  block updates  until convergence.
In particular, the number of blocks in each iteration of the element-wise BCD algorithm is $\frac{M+1}{2}$ times higher than that of the AO-MM algorithm.
On the other hand, as globally optimal solutions \eqref{eq11} are obtained for all the blocks of the element-wise BCD algorithm, the number of iterations needed for convergence is less than for the AO-MM algorithm, in which the phase shifts are updated in parallel but sub-optimally by \eqref{eq21}.
Therefore, there is a trade-off between the number of blocks per iteration and the number of iterations required for convergence, which shall be investigated in the next section.

\section{Simulation Results}
In this section, we numerically evaluate the performance of the proposed algorithms. The channels are assumed to be independent Rayleigh fading, and the path loss exponent is denoted by $\alpha$ with reference distance 10 meters. The noise power at both the legitimate receiver and the eavesdropper is set to  $\sigma_\mathrm{l}^2=\sigma_\mathrm{e}^2=-80$ dBm. The distance between the transmitter and the IRS is denoted as $r_\mathrm{TR}$, while $r_\mathrm{Rl}$ and $r_\mathrm{Re}$ are the distances from the IRS to the legitimate receiver and the eavesdropper, respectively. The simulation results in Figs. \ref{fig1} and \ref{fig2} are averaged over 1000 channel realizations.

\begin{figure}[t]
	\centering\includegraphics[height=6cm]{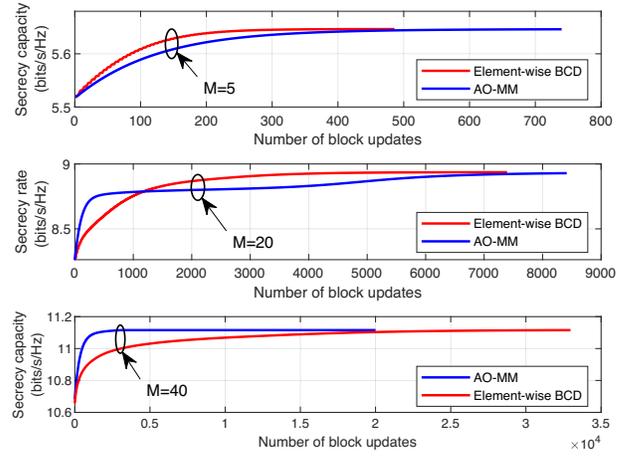}
	\caption{Convergence of the proposed algorithms for different values of $M$ when $\Nt=5$, $\alpha=4$,  $P=5$ dBm, $r_\mathrm{TR}=250$ m, $r_\mathrm{Rl}=r_\mathrm{Re}=160$ m.}\label{fig0}
\end{figure}
\subsection{Comparison of the Proposed Algorithms}
The convergence of the proposed algorithms is investigated for three typical examples in Fig. \ref{fig0}. 
The stopping criterion for convergence is that the increment of the normalized objective function is less than $\epsilon=10^{-6}$.
We first investigate the scenario where the thin film employed to implement the IRS has a small area, e.g., $M=5$. Although the number of blocks per iteration of the element-wise BCD algorithm is slightly larger for the AO-MM algorithm (6 versus 2), the element-wise BCD algorithm needs much fewer iterations for convergence since globally optimal solutions are obtained for all the blocks. Therefore, the element-wise BCD algorithm converges faster when $M$ is relatively small.
On the other hand, as the value of $M$ gradually increases, i.e., $M=40$, a large number of blocks (41 blocks) have to be updated in each iteration of the element-wise BCD algorithm, which slows down the convergence. In contrast, only two blocks per iteration have to be updated in the AO-MM algorithm. This offsets the drawback that the AO-MM requires more iterations due to the sub-optimal solutions for each phase shift update block. In summary, the element-wise BCD algorithm is suitable for small-scale IRS systems while the AO-MM algorithm is preferable for large-scale IRS systems.

\subsection{Average Secrecy Rate Evaluation}

\begin{figure}[t]
	\centering\includegraphics[height=6cm]{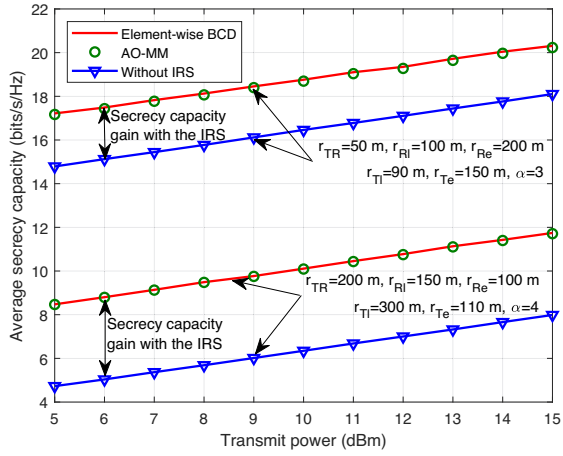}
	\caption{Average secrecy rate achieved by different algorithms when $M=10$ and $\Nt=8$.}\label{fig1}
\end{figure}

In Fig. \ref{fig1}, the average achievable secrecy rate is plotted for different algorithms. 
First, we observe that the average secrecy rate achieved by both proposed algorithms is the same.
We also compare our approach with a benchmark system which does not employ an IRS for security provisioning.
In this case, the distance between the transmitter and the legitimate receiver is denoted by $r_\mathrm{Tl}$ while the distance between the transmitter and the eavesdropper is denoted by $r_\mathrm{Te}$\footnote{
The values of $r_\mathrm{Tl}$ and $r_\mathrm{Te}$ and the  performance gains achievable with IRSs depend on the geometry of the network.
In Fig. \ref{fig1}, we investigate different geometries by providing one example with $r_\mathrm{Tl}<r_\mathrm{Rl}$ and $r_\mathrm{Te}<r_\mathrm{Re}$, and one example with $r_\mathrm{Tl}>r_\mathrm{Rl}$ and $r_\mathrm{Te}>r_\mathrm{Re}$.
}. To maximize the secrecy rate, optimal transmit beamforming according to \eqref{eq4} is adopted. 
As can be observed from Fig. \ref{fig1}, the system with IRS provides a significant performance gain in terms of the secrecy rate, which indicates that deploying IRSs is a promising approach for improving the physical layer security of wireless communications systems.
\subsection{Massive MIMO or Massive IRS?}
\begin{figure}[t]
	\centering\includegraphics[height=6cm]{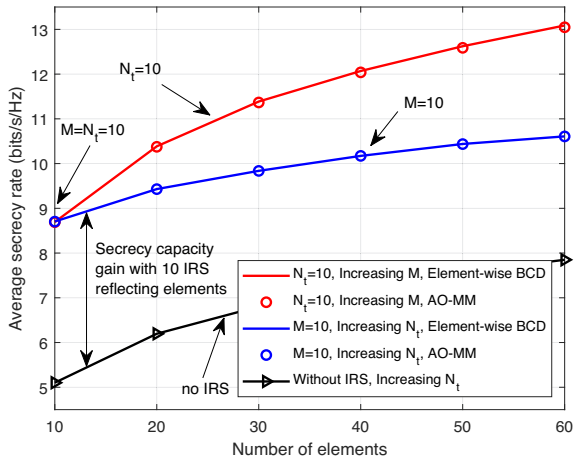}
	\caption{Average secrecy rate achieved for different values of  $M$ and $\Nt$, when $P=5$ dBm, $\alpha=4$, $r_\mathrm{TR}=200$ m, $r_\mathrm{Rl}=150$ m, $r_\mathrm{Re}=100$ m, $r_\mathrm{Tl}=300$ m, and $r_\mathrm{Te}=110$ m.}\label{fig2}
\end{figure}
For conventional wireless communications systems,  deploying large-scale antenna arrays at the transceivers is an effective way to boost communication performance, including the network capacity and physical layer security. Therefore, it is intriguing to investigate how much performance gain we can obtain from large-scale IRSs. In Fig. \ref{fig2}, we first increase the number of reflecting elements of the IRS while keeping the number of antenna elements as $\Nt=10$ (red curve). In addition, to illustrate the effectiveness of the IRS, we also evaluate the average secrecy rate achieved for different numbers of transmit antenna elements when using an IRS with $M=10$ (blue curve). As can be observed from Fig. \ref{fig2}, increasing the number of IRS reflecting elements is more beneficial for improving the secrecy rate than increasing the number of antenna elements. Moreover, as the IRS is a passive device, deploying large-scale IRSs is more energy-efficient than installing more energy-consuming RF chains and power amplifiers as is needed for increasing the number of antenna elements at the transmitter.
These results clearly demonstrate the superiority of IRSs compared with conventional system designs in terms of both communication performance and energy consumption.

\section{Conclusions}
In this paper, we proposed to improve the physical layer security of wireless communications networks by deploying IRSs. Two efficient algorithms, i.e., the element-wise BCD and AO-MM algorithms, were developed for joint optimization of the beamformer at the transmitter and the phase shifts at the IRS. The element-wise BCD algorithm was shown to be preferable for small-scale IRS-assisted systems, while the AO-MM algorithm is advantageous for wireless systems with large-scale IRSs. Simulation results have confirmed the huge potential of IRSs to improve the security and energy efficiency of future communications systems.

\appendix

\section{}\label{appA}
As the phase shift matrix $\mathbf{\Phi}$ is a diagonal matrix, the objective function of $\mathcal{P}_3$ can be rewritten as a function of the $k$-th reflecting element as follows:
\begin{equation}
\frac{1+\frac{1}{\sigma_\mathrm{l}^2}\left|\mathbf{h}_\mathrm{l}^H\mathbf{\Phi}\mathbf{G}\mathbf{f}\right|^2}{1+\frac{1}{\sigma_\mathrm{e}^2}\left|\mathbf{h}_\mathrm{e}^H\mathbf{\Phi}\mathbf{G}\mathbf{f}\right|^2}
=\frac{c_{\mathrm{l},k}+d_{\mathrm{l},k}\cos(\theta_k+p_{\mathrm{l},k})}{c_{\mathrm{e},k}+d_{\mathrm{e},k}\cos(\theta_k+p_{\mathrm{e},k})},
\end{equation}
where $c_{i,k}$, $d_{i,k}$, and $p_{i,k}$ are given in \eqref{eq11}.
By taking the derivative of the objective function with respect to $\theta_k$, and setting the derivative to zero, we obtain the following equation
\begin{equation}
\begin{split}
d_{\mathrm{l},k}\sin(\theta_k+p_{\mathrm{l},k})\left[c_{\mathrm{e},k}+d_{\mathrm{e},k}\cos(\theta_k+p_{\mathrm{e},k})\right]\\
=d_{\mathrm{e},k}\sin(\theta_k+p_{\mathrm{e},k})\left[c_{\mathrm{l},k}+d_{\mathrm{l},k}\cos(\theta_k+p_{\mathrm{l},k})\right].
\end{split}
\end{equation}
This equation can be further simplified by some basic trigonometric manipulations as follows,
\begin{equation}
A_k\sin\theta_k+B_k\cos\theta_k=d_{\mathrm{l},k}d_{\mathrm{e},k}\sin(p_{\mathrm{e},k}-p_{\mathrm{l},k}),
\end{equation}
where
\begin{equation}
\begin{split}
A_k\triangleq c_{\mathrm{e},k}d_{\mathrm{l},k}\cos p_{\mathrm{l},k}-c_{\mathrm{l},k}d_{\mathrm{e},k}\cos p_{\mathrm{e},k},\\
B_k\triangleq c_{\mathrm{e},k}d_{\mathrm{l},k}\sin p_{\mathrm{l},k}-c_{\mathrm{l},k}d_{\mathrm{e},k}\sin p_{\mathrm{e},k}.
\end{split}
\end{equation}
When $A_k\ge0$, by introducing an auxiliary angle, the equation can be recast as
\begin{equation}
\cos\left(\theta_k-\arctan\frac{A_k}{B_k}\right)=\frac{d_{\mathrm{l},k}d_{\mathrm{e},k}\sin(p_{\mathrm{e},k}-p_{\mathrm{l},k})}{\sqrt{A^2+B^2}}.
\end{equation}
It can be readily shown by checking the second derivative that the objective function is maximized when
\begin{equation}
\theta_k=\arctan\frac{A_k}{B_k}-\arccos\frac{d_{\mathrm{l},k}d_{\mathrm{e},k}\sin(p_{\mathrm{e},k}-p_{\mathrm{l},k})}{\sqrt{A^2+B^2}}.
\end{equation}
The optimal solution when  $A_k<0$ can be obtained in a similar manner, which completes the proof.


\bibliographystyle{IEEEtran}
\bibliography{bare_conf}

\end{document}